\documentclass[reprint,superscriptaddress,pra]{revtex4}
\usepackage{url}
\usepackage{amsmath}
\usepackage{mathtools}
\usepackage{bbold}
\usepackage{mathrsfs}

\usepackage{graphicx}
\usepackage[caption=false]{subfig}
\usepackage{epstopdf} 
\usepackage{wrapfig}
\usepackage{braket}
\usepackage{xcolor}
\usepackage{comment}
\usepackage{lipsum}
\usepackage{cancel}
\usepackage{hyperref}

\usepackage{tikz}
\usepackage{circuitikz}
\usetikzlibrary{chains,fit,shapes}
\usetikzlibrary{cd} 
\usetikzlibrary{shapes.geometric}
\usetikzlibrary{decorations.markings}
\usetikzlibrary{decorations.pathmorphing,patterns}
\usetikzlibrary{decorations}
\usetikzlibrary{calc}
\usetikzlibrary{intersections}
\usetikzlibrary{arrows}
\usetikzlibrary{matrix}
\usetikzlibrary{positioning}
\usetikzlibrary{automata}

\newcommand{\RNum}[1]{\uppercase\expandafter{\romannumeral #1\relax}}

\newcommand{\btheta}{\boldsymbol{\theta}}

\def\ket#1{| #1\rangle}
\newcommand{\nep}{\mathrm{e}}

\newcommand{\upket}{|\uparrow\rangle}
\newcommand{\downket}{|\downarrow\rangle}

\newcommand{\x}{{\mathbf x}}

\newcommand{\PauliSigma}{\hat{\sigma}}

\newcommand{\identity}{\mathbf{1}}

\newcommand{\Ho}{\hat{H}}

\newcommand{\Nsites}{\mathrm{N}}

\newcommand{\Ptrot}{\mathrm{P}}

\newcommand{\gs}{\mathrm{\scriptstyle gs}}

\graphicspath{{./}{./Fig/}}

\begin{document}
\author{Ruiyi Wang$^*$}
\affiliation{SISSA, Via Bonomea 265, I-34136 Trieste, Italy}

\author{Anita Weidinger$^*$}
\affiliation{Institute for Theoretical Physics, University of Innsbruck, A-6020 Innsbruck, Austria}
\affiliation{Parity Quantum Computing GmbH, A-6020 Innsbruck, Austria}

\author{Glen Bigan Mbeng}
\affiliation{Institute for Theoretical Physics, University of Innsbruck, A-6020 Innsbruck, Austria}
\affiliation{Parity Quantum Computing GmbH, A-6020 Innsbruck, Austria}

\author{Wolfgang Lechner}
\affiliation{Institute for Theoretical Physics, University of Innsbruck, A-6020 Innsbruck, Austria}
\affiliation{Parity Quantum Computing GmbH, A-6020 Innsbruck, Austria}
\affiliation{Parity Quantum Computing Germany GmbH, 20095 Hamburg, Germany}

\author{Giuseppe E. Santoro}
\affiliation{SISSA, Via Bonomea 265, I-34136 Trieste, Italy}
\affiliation{International Centre for Theoretical Physics (ICTP), P.O.Box 586, I-34014 Trieste, Italy}


\title{Parity Mapping for Quantum Optimization on Frustrated Ising Rings}

\begin{abstract}
The frustrated Ising ring is one of the simplest models exhibiting exponential closing spectral gaps, making it a paradigmatic and challenging benchmark for quantum annealing (QA). 
Ground-state preparation for this model has therefore been studied extensively in both continuous-time QA and digitized protocols such as the Quantum Approximate Optimization Algorithm (QAOA). 
Here, we use the frustrated Ising ring to investigate how the parity mapping affects the performance of both QA and QAOA. For QA, finite-size calculations show that the parity mapping increases the minimum spectral gap under the energy normalization used in this work, thereby enabling faster continuous-time ground state preparation protocols. An ideal implementation of Parity-QA, with a single global constraint, shows no evidence of exponential gap closing over the accessible system sizes, whereas a hardware-motivated decomposition into local constraints restores the exponential decrease, albeit with a smaller fitted exponent than conventional QA. For the digitized protocol, we find that the number of Parity-QAOA layers required to prepare the exact ground state remains constant over the simulated sizes, improving upon the quadratic scaling required by conventional QAOA.
To investigate the role of constraints in Parity-QAOA, we further consider a modified Ising ring instance in which the constraint term is essential for preparing the target ground state. We then compare the corresponding resource requirements with those of conventional QAOA.
\end{abstract}

\maketitle
\def\thefootnote{*}\footnotetext{These authors contributed equally to this work}\def\thefootnote{\arabic{footnote}}

\section{Introduction}\label{sec:introduction}
Quantum computing has seen substantial progress in both hardware and software over the past years, enabling increasingly sophisticated investigations of scientific and computational problems~\cite{Nielsen_Chuang:book,Montanaro_npj2016,Alexeev_PRXQuantum2021}.
Translating these advances into a practically relevant computational advantage, however, requires identifying algorithms that use quantum resources efficiently while remaining compatible with the connectivity, coherence, and noise constraints of current hardware~\cite{preskill2018quantum,Bharti_RMP2022}. Quantum optimization provides a prominent setting in which to address this challenge.

A well-established continuous-time approach to quantum optimization is quantum annealing (QA), which aims to prepare the ground state of a target Hamiltonian $\Ho_\mathrm{targ}$ encoding the solution to a potentially hard optimization problem~\cite{finnila_quantum_1994,kadowaki1998quantum,Santoro_SCI02,Santoro_2006,Albash_RMP18}. In QA, the system is initialized in the ground state of a driving (or mixer) Hamiltonian, usually chosen to be $\Ho_{\mathrm{drive}}=\Ho_x =\sum_j^N \PauliSigma_j^x$, where the pauli operators $\PauliSigma_j^x$ act on each of the $N$ qubits, and subsequently evolved for a total annealing time $\tau$ under a time-dependent Hamiltonian $\Ho(t)$ that interpolates between $\Ho(0)=\Ho_{\mathrm{drive}}$ and $\Ho(\tau)=\Ho_\mathrm{targ}$. The  standard QA interpolating Hamiltonian is:
\begin{equation}
     \Ho(t) = s(t) \, \Ho_\mathrm{targ} + (1-s(t)) \, \Ho_\mathrm{drive} \,,
\label{eqn:ann_hamiltonian_s}
\end{equation}
where the schedule function $s(t)$ satisfies $s(0)=0$ and $s(\tau)=1$. QA's success usually relies on the adiabatic theorem which ensures that, if the evolution is sufficiently slow relative to the instantaneous spectral gaps, the system remains in the ground state of the instantaneous Hamiltonian throughout the dynamics~\cite{Jansen_JMP2007, 
Messiah_Book2014, Albash_RMP18}. Consequently, a minimum gap that closes exponentially with system size can make conventional QA with a linear schedule, $s(t)=t/\tau$, impractical by requiring exponentially long annealing times to maintain adiabaticity. Optimized schedules and non-adiabatic protocols can mitigate or circumvent such bottlenecks; however, their performance and practical relevance depend on the available controls and the cost of optimizing the protocol~\cite{Roland_PRA2002,Cote_2023,balducci2024fighting}.

The quantum approximate optimization algorithm (QAOA) provides a digital alternative to QA based on alternating unitary evolutions generated by the target and driving Hamiltonians~\cite{farhi_quantum_2014,zhou_quantum_2020,Blekos_PhyRep2024}. Indeed, QAOA belongs to the broader class of variational quantum algorithms that hold the potential to overcome the small-gap bottleneck encountered in QA~\cite{cerezo_variational_2021}. 
More specifically, in QAOA, the system is initialized in the ground state of the driver Hamiltonian and subsequently evolved through a sequence of parametrized quantum gates until a final state is reached. 
These gates correspond to alternating unitary transformations generated by the target and driver Hamiltonians. The parameters in these gates are then optimized classically to minimize the expectation value of the target Hamiltonian evaluated on the final quantum state. When the optimization is successful, the resulting state approximates, or in some cases exactly reproduces, a ground state of the target Hamiltonian. Unlike adiabatic QA, finite-depth QAOA does not require the system to follow an instantaneous eigenstate and is therefore not directly constrained by the minimum gap of a prescribed interpolation. Instead, in the non-adiabatic regime, QAOA's output state can be naturally viewed as the solution to a discrete-time quantum optimal control problem~\cite{Dalessandro2007,Yang_PRX2017,mbeng2019optimal,Brady_PRL2021} and its performance is governed by the structure of the reachable manifold and the trainability of the variational landscape~\cite{arezzo2025digital,wang2025exponential}. 
In particular, as argued in Ref.~\cite{lloyd2014information}, the minimum number of control parameters required to prepare the target ground state with finite accuracy is determined by the dimension of the reachable manifold.

A major challenge in implementing QA and QAOA efficiently on current noisy intermediate-scale quantum (NISQ) devices~\cite{preskill2018quantum} is the limited connectivity of the physical hardware. Target Hamiltonians may involve all-to-all or long-range interactions, whereas most hardware platforms typically support only quasi-local couplings. 
The Lechner–Hauke–Zoller (LHZ), or Parity, architecture addresses this challenge by mapping interacting pairs of logical spins to parity qubits that encode their relative alignment~\cite{lechner2015quantum,Ender_Quantum2023}. After this Parity mapping, logical interactions are encoded in fully programmable local fields on parity qubits, at the cost of introducing parity constraints. These constraints are then enforced by penalty terms in continuous-time algorithms and by additional constraint unitaries in digital algorithms. Implementations of QA and QAOA within the parity architecture, referred to as Parity-QA~\cite{lechner2015quantum} and Parity-QAOA \cite{lechner2020quantum}, have demonstrated advantages across various models \cite{Susa_PRA2021, Lanthaler_PRL2023,weidinger2023error,weidinger2024performance}. 
The Parity mapping restructures the accessible Hilbert space and introduces global constraints that must be enforced either through nonlocal penalty terms (in QA) or additional constraint unitaries (in QAOA). 
While enabling more efficient and less error-prone simulations on current devices, this approach typically involves trade-offs such as increased qubit overhead and a greater number of QAOA unitaries on the tested cases.

Previous studies of Parity-QA and Parity-QAOA focused on highly connected target Hamiltonians for which the parity mapping introduces a substantial qubit overhead~\cite{Lanthaler_PRL2023,Fellner_Quantum2023,Wybo_Quantum2024,Bennett_PRA2025}. In this work, we investigate parity-based optimization for a sparse target Hamiltonian, namely the frustrated Ising ring model and its variants~\cite{Knysh_PRA2020}. These models are of interest both for theoretical studies of QA and QAOA, and for efficient practical implementations within the parity architecture. 
The frustrated Ising ring, proposed in Ref.~\cite{Knysh_PRA2020}, is one of the simplest models exhibiting exponentially closing spectral gaps, making it a compelling test case for continuous-time approaches that extend traditional QA \cite{Cote_2023,balducci2024fighting}. 
In Refs.~\cite{wang2025exponential,arezzo2025digital}, digital QAOA approaches applied to this model demonstrated a remarkable exponential speedup with respect to standard QA, requiring a number of alternating unitaries that only grows quadratically with system size for exact ground-state preparation. 
Moving towards the parity framework, we find that these ring models differ from previously studied highly connected models because they incur zero qubit overhead and require fewer unitaries than standard QAOA. They therefore provide valuable test cases and serve as a natural starting point for evaluating parity-based methods on sparse graphs. 
Moreover, we observe an amplification of the spectral gaps when using Parity-QA, compared to standard linear-schedule QA, which suggests the possibility of mitigating the exponentially closing gap under ideal annealing hardware. The frustrated ring model also represents a special case for Parity-QAOA, as it is the first instance in which the unitary associated with the Parity-constraint is not required, and the problem can be solved using only single-qubit rotations. 
Finally, to further clarify the role of the constraint, we examine additional variants of the model.

The manuscript is structured as follows. 
Section \ref{sec:parity_mapping} introduces the Parity transformation on the Hamiltonians of interest. 
Section \ref{sec:Parity-QAOA} introduces the original QAOA and its parity-adapted version, which we call henceforth Parity-QAOA. 
Section \ref{sec:models} presents the frustrated Ising ring model in its original formulation, as well as a few variants that we will study.  
We further report numerical results on the scaling of the minimal spectral gap in Parity-QA as compared with standard QA. 
Section \ref{sec:results} presents the results obtained using Parity-QAOA. In Section \ref{sec:improved controllability}, we analyze the case in which the protocol omits the constraint and achieves the exact solution with a constant resource scaling independent of system size. Finally, in Sec.~\ref{sec:use_of_constraint}, we examine the case where the constraint is included and compare the resulting CNOT depth and CNOT gate count with those of standard QAOA. 

\section{The Parity Transformation} \label{sec:parity_mapping}

\begin{figure}[htp]
\centering
\includegraphics[width=1\columnwidth]
{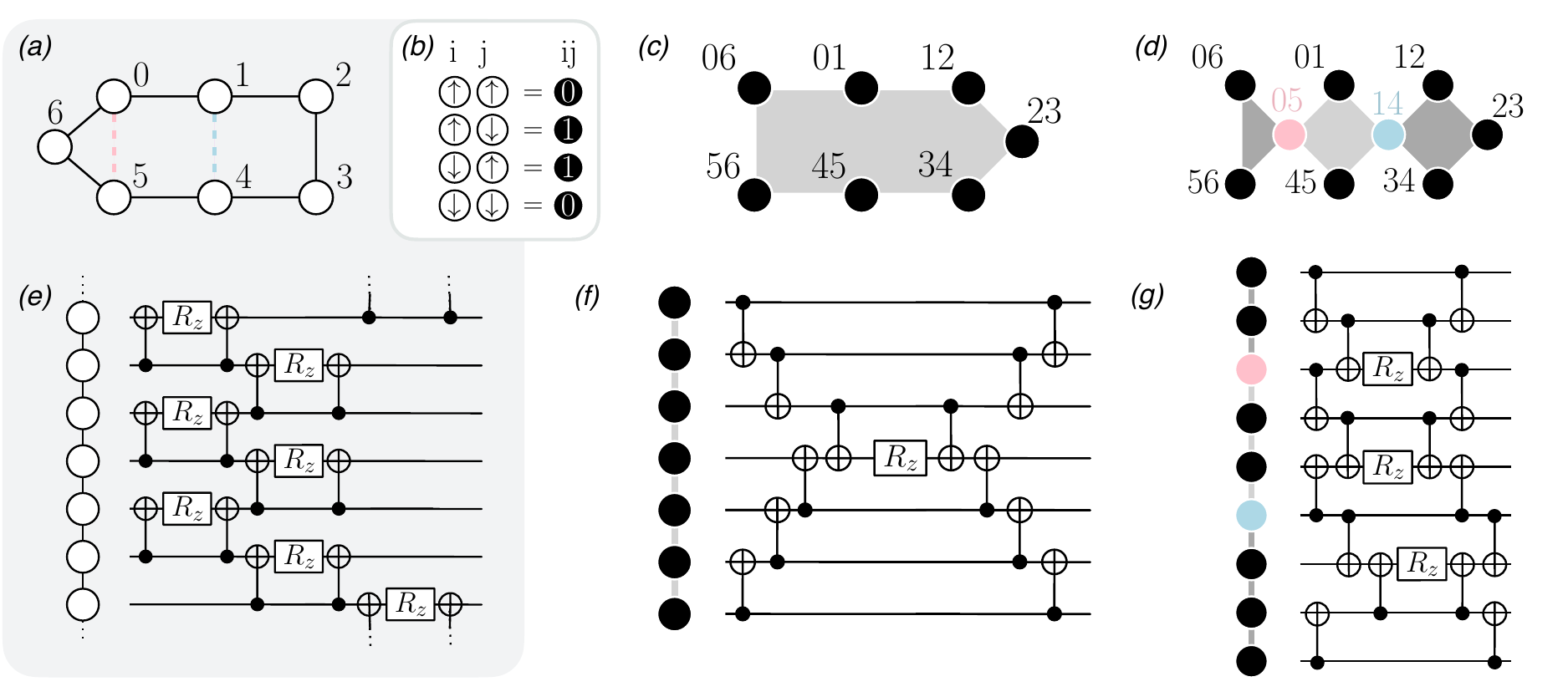}
\caption{The Ising ring with $\Nsites=7$ in its logical representation (a) and parity representations (c)–(d). The parity transformation table is given in (b). Without ancillas the parity graph has an $N$-body constraint visualized by a grey polygon in (c). With ancilla interactions (red and blue dashed lines in (a)) the parity graph has three constraints of size three or four, shown in (d). The digital implementations of the multi-qubit gates are shown in (e), (f) and (g) for the logical graph, parity graph and the parity graph with ancillas, respectively.}
\label{fig_notes:mapping}
\end{figure}

In this section, we discuss the main technique in this work, namely the Parity or LHZ architecture~\cite{lechner2015quantum}. 
We consider the problem of preparing the ground state of the following Ising ring target Hamiltonian~\cite{Mbeng_SciPost2024}
\begin{equation} \label{eqn:H_targ}
    \Ho_\mathrm{targ} = \Ho_z = -\sum_{j=1}^\Nsites J_j \PauliSigma^z_j\PauliSigma^z_{j+1} \;,
\end{equation}
with periodic boundary conditions $\PauliSigma^z_{\Nsites+1}=\PauliSigma_1^z$. An example of a ring with $N=7$ is shown in Fig.~\ref{fig_notes:mapping} (a).

The parity mapping $\PauliSigma^z_j \PauliSigma^z_{j+1} \rightarrow \tilde{\sigma}_k^z$ maps pairs of interacting logical qubits to individual physical (or parity) qubits. The translation table is shown in Fig.~\ref{fig_notes:mapping} (b). If two logical qubits $\PauliSigma^z_j$ and $ \PauliSigma^z_{j+1}$ are parallel or anti-parallel the corresponding $k$-th parity qubit is $0$ or $1$, respectively.
After the Parity mapping, the Hamiltonian $\Ho_z$ encoding the Ising ring model reads:
\begin{equation}
    \Ho_z \rightarrow \tilde{H}_z = -\sum_{k=1}^{K}\tilde{J}_{k} \tilde{\sigma}^z_k,
\end{equation}
with $\tilde{J}_k=J_k$ and $K=\Nsites$: 
The number of physical qubits $K$ coincides with the number of pairwise interactions in the target Hamiltonian, and hence in this case equals $\Nsites$, the number of logical qubits in the ring structure. 

The full $2^K$-dimensional Hilbert space of the parity qubits contains configurations that cannot be decoded into logical spin assignments. For example, for odd $N$, the physical configuration $ |1\rangle^{\otimes K}$ cannot be decoded to a logical one because no logical state has all neighboring qubits anti-parallel to each other. Therefore,  constraints are required to guarantee the logical validity of physical configurations.
In Ref.~\cite{lechner2015quantum}, the number of constraints needed is specified to be $K-\Nsites+1$. Hence, a single constraint is required for the parity transformation of the Ising ring model at hand~\footnote{Here $K=\Nsites$, but one parity configuration can be decoded into two logical ones and, therefore, $2^{\Nsites-1}$ physical states correspond to $2^\Nsites$ logical ones, the remainder are non-valid states. Hence, one constraint is needed.}.
In particular, a valid state $\ket{\tilde{\psi}_{\mathrm{valid}}}$ satisfies the following identity:
\begin{equation}
    \prod_{k=1}^K \tilde{\sigma}^z_k \ket{\tilde{\psi}_{\mathrm{valid}}}= (\PauliSigma^z_1 \PauliSigma^z_2) (\PauliSigma^z_2 \PauliSigma^z_3) \cdots  (\PauliSigma^z_{K-1} \PauliSigma^z_{K})( \PauliSigma^z_{K} \PauliSigma^z_{1})\ket{\tilde{\psi}_{\mathrm{valid}}} = \ket{\tilde{\psi}_{\mathrm{valid}}}.
\end{equation}
In other words, a valid computational-basis configuration contains an even number of parity qubits occupying the state $\ket{1}$. 
To enforce this constraint, we add a non-local penalty term to the target Hamiltonian, denoted as $\tilde{H}_c$:
\begin{equation}\label{eq:constraint_Hamiltonian}
    \tilde{H}_c = -\prod_{k=1}^K \tilde{\sigma}^z_k \,,
\end{equation}
%
which is minimized by all valid states.

To solve the optimization problem within the parity picture, we must minimize the following  target Parity Hamiltonian:
\begin{equation} \label{eqn:H_targ_parity}
    \tilde{H}_\mathrm{targ} = \tilde{H}_z + C \tilde{H}_c \;.
\end{equation}
Here, $C$ is a constraint strength that should be chosen sufficiently large to ensure that the ground state of $\tilde{H}_\mathrm{targ}$ is a valid state.  Ref.~\cite{Lanthaler_2021NJP} investigates the minimal required constraint strength and its scaling for various problem classes. For the problem we study in this work, the Ising ring, the constraint strength should be $C>\ \text{min}_{j}|J_{j}|$. 
Finally, the driving Hamiltonian in Parity-QA and Parity-QAOA can be defined as the driving Hamiltonian defined in the Introduction, Sec.~\ref{sec:introduction}, namely $\tilde{H}_\mathrm{drive} = \tilde{H}_x = -\sum_{k=1}^K \tilde{\sigma}^x_k$. Here, the $\tilde{\sigma}^x_k$ operators now act on $K$ parity qubits.

Fig.~\ref{fig_notes:mapping} (c) illustrates the parity transformation of the Ising ring depicted in Fig.~\ref{fig_notes:mapping} (a). The constraint $\tilde{H}_c$, which is visualized as a gray area, involves a K-qubit term that is not practical for current annealing hardware, especially as the system size grows. 
To address this issue, ancilla qubits can be introduced to break the K-qubit term into 3- and 4-qubit local terms. These ancilla qubits usually correspond to pairs of non-interacting logical qubits in the original problem. An example is given in Fig.~\ref{fig_notes:mapping}~(a), where the colored dashed lines (red and blue) indicate auxiliary interactions. The corresponding parity architecture is shown in Fig.~\ref{fig_notes:mapping}~(d). The constraint Hamiltonian of Eq.~\eqref{eq:constraint_Hamiltonian} with ancilla qubits, as discussed in Ref.~\cite{Ender_Quantum2023}, changes to $\tilde{H}_c = -\sum_{l=1}^{L}\prod_{k \in C_{l}} \tilde{\sigma}^z_{k}$ where $L$ is the total number of constraints and $C_{l}$ is the set of qubit indices participating in the l-th constraint.
The ancilla qubits play a crucial role in enabling implementations of quantum annealing on analog NISQ devices. 
Therefore, as we study the spectra in QA and Parity-QA in Section \ref{sec:min_gap}, we also explicitly investigate how these ancilla qubits affect the spectral gaps.

\section{Parity-QAOA} \label{sec:Parity-QAOA}
The Quantum Approximate Optimization Algorithm (QAOA), later referred to as standard QAOA, adopts an alternating {\em Ansatz} in evolving the quantum states. 
The final state $\ket{\psi_\Ptrot(\btheta^x,\btheta^z)}$ is obtained from 
\begin{equation}
\ket{\psi_\Ptrot(\btheta^x,\btheta^z)} =  \hat{U}_x(\theta^x_\Ptrot)\, \hat{U}_z(\theta^z_\Ptrot) \cdots \hat{U}_x(\theta^x_1)\, \hat{U}_z(\theta^z_1)\ket{\psi_0}
\end{equation}
where the initial state  $\ket{\psi_0} = |+\rangle^{\otimes \Nsites} = \left( \frac{\upket+\downket}{\sqrt{2}}  \right)^{\otimes \Nsites}$. The alternating unitary operators are defined using the driving Hamiltonian and the target Hamiltonian, respectively. At each layer $p = 1, \cdots, \Ptrot$, we have:
\begin{equation}
\hat{U}_x(\theta^x_p) = \nep^{-i\theta^x_p \hat{H}_x} \;, \hspace{10mm} 
\hat{U}_z(\theta^z_p) = \nep^{-i\theta^z_p \hat{H}_{\mathrm{targ}}} = \nep^{-i\theta^z_p \hat{H}_z} \;.
\end{equation}
The $2\Ptrot$ parameters $(\btheta^x,\btheta^z)$ are optimized by minimizing the variational energy:
\begin{equation} \label{eqn:E_QAOA}
E_{\Ptrot}(\btheta^x,\btheta^z) = \langle \psi_{\Ptrot}(\btheta^x,\btheta^z)| \Ho_\mathrm{targ} | \psi_{\Ptrot}(\btheta^x,\btheta^z)\rangle  \;.
\end{equation}

A similar {\em Ansatz} can be written for the Parity-QAOA~\cite{lechner2020quantum}, following the parity mapping of Sec.~\ref{sec:parity_mapping}: 
\begin{equation} 
\label{eqn:parity_QAOA_ansatz}
|\tilde{\psi}_{\Ptrot}(\btheta^x,\btheta^c,\btheta^z)\rangle =
\tilde{U}_x(\theta^x_\Ptrot)\, \tilde{U}_c(\theta^c_\Ptrot)\, \tilde{U}_z(\theta^z_\Ptrot)\cdots \tilde{U}_x(\theta^x_1)\, \tilde{U}_c(\theta^c_1)\, \tilde{U}_z(\theta^z_1)
\, |\tilde{\psi}_0\rangle \;.
\end{equation}
The initial state is $|\tilde{\psi}_0\rangle  = \ket{+}^{\otimes K}$.  Three types of unitaries are applied to the quantum state at each layer $p$, and this is due to the additional constraint term $\tilde{H}_c$ in the target Hamiltonian $\tilde{H}_\mathrm{targ}$. 
These unitaries read:
\begin{equation}
\tilde{U}_x(\theta^x_p) = 
\nep^{-i\theta^x_p\tilde{H}_x} \;, \hspace{10mm}
\tilde{U}_c({\theta^c_p}) = 
\nep^{-i\theta^c_p \tilde{H}_c} \;, \hspace{10mm}
\tilde{U}_z(\theta^z_p) = \nep^{-i\theta^z_p\tilde{H}_z} \;.
\end{equation}
At each layer, there are three free parameters $(\theta^x_p,\theta^c_p, \theta^z_p)$ and hence the final state at layer $\Ptrot$ is determined by $3\Ptrot$ parameters that are determined by an external classical optimization loop. The variational energy now reads:
\begin{equation} \label{eqn:E_parity_QAOA}
E_{\Ptrot}(\btheta^x,\btheta^c,\btheta^z) = \langle \tilde{\psi}_{\Ptrot}(\btheta^x,\btheta^c,\btheta^z)| \tilde{H}_\mathrm{targ}  | \tilde{\psi}_{\Ptrot}(\btheta^x,\btheta^c,\btheta^z)\rangle  \;,
\end{equation}
where $\tilde{H}_\mathrm{targ}$ is as defined in Eq.~\eqref{eqn:H_targ_parity}.

In the standard QAOA, the interaction unitary $\hat{U}_z(\theta^z_p)=\prod_{j=1}^N\nep^{-i\theta^z_p J_j\PauliSigma^z_j\PauliSigma^z_{j+1}}$ can be compiled with a single-qubit rotation $R^{(j)}_z(\alpha)=\nep^{-i\alpha\PauliSigma_j^z/2}$ and two controlled-NOT (CNOT) gates for each edge $(j,j+1)$ of the ring, using the identity
$
\nep^{-i\alpha\PauliSigma^z_j\PauliSigma^z_{j+1}}=
\operatorname{CNOT}_{j,j+1} R^{(j+1)}_z(2\alpha)\operatorname{CNOT}_{j,j+1}
$. The interaction unitary $\hat{U}_z(\theta)$ therefore uses $2\Nsites$ CNOT gates per layer. Assuming a ring connectivity for the quantum device and scheduling the ring edges in parallel gives CNOT depth four for even $\Nsites$ and six for odd $\Nsites$, as illustrated in Fig.~\ref{fig_notes:mapping}(e).

In the Parity-QAOA implementation, the unitaries $\tilde{U}_x(\theta^x_p) $ and $\tilde{U}_z(\theta^z_p)$ require only single qubit rotations, while the constraint unitary 
%
%
$\tilde{U}_c(\theta^c_p)$ involves a $K$-qubit entangling gate.
It can be implemented with the circuit shown in Fig.~\ref{fig_notes:mapping} (f), comprising a chain sequence of CNOT gates connecting all physical qubits, followed by a single qubit z-rotation $R_z$, followed by the reversed preceding CNOT chain.
The number of CNOT gates required to implement the constraint unitary $\tilde{U}_c$ is $2(\Nsites-1)$, and the CNOT depth is $\Nsites$ for even $\Nsites$ and $\Nsites+1$ for odd $\Nsites$ (see Fig.~\ref{fig_notes:mapping}, f).

Thus, per QAOA layer, standard QAOA has constant CNOT depth and $2\Nsites$ CNOTs, whereas the global-constraint Parity-QAOA implementation has linear CNOT depth and $2(\Nsites-1)$ CNOTs. Whether this trade-off is advantageous depends on the number of layers required, the physical connectivity, and the relative importance of entangling-gate errors and circuit duration.

\section{Frustrated Ising ring models} \label{sec:models}
\begin{figure}[!htp]
\centering
\begin{tikzpicture}[scale=1.7,line cap=round,line width=2pt]
\filldraw [fill=black!0!white] (0,0) circle (2cm);
\draw [line width = 0.3mm, draw=blue, dashed] (-2.2,0) -- (2.0,0) node[right, black] {};
\foreach \x [evaluate=\x as \angle using (\x-0.5)*360/13] in {1,...,13}
{
\draw[line width=1pt,fill=white] ({2*cos(\angle)},{2*sin(\angle)}) circle (3mm);
}
\node[draw=none,font=\tiny,text=black,scale=1.5] at ({2*cos(0.5*360/13)},{2*sin(0.5*360/13)}) {$1$};
\node[draw=none,font=\tiny,text=black,scale=1.5] at ({2*cos(1.5*360/13)},{2*sin(1.5*360/13)}) {$2$};
\node[draw=none,font=\tiny,text=black,scale=1.5] at ({2*cos(2.5*360/13)},{2*sin(2.5*360/13)}) {$\cdots$};
\node[draw=none,font=\tiny,text=black,scale=1.5] at ({2*cos(4.5*360/13)},{2*sin(4.5*360/13)}) {$\cdots$};
\node[draw=none,font=\tiny,text=black,scale=1.5] at ({2*cos(5.5*360/13)},{2*sin(5.5*360/13)}) {$\scriptstyle{\frac{\Nsites-1}{2}}$};
\node[draw=none,font=\tiny,text=black,scale=1.5] at ({2*cos(6.5*360/13)},{2*sin(6.5*360/13)}) {$\scriptstyle{\frac{\Nsites+1}{2}}$};
\node[draw=none,font=\tiny,text=black,scale=1.5] at ({2*cos(7.5*360/13)},{2*sin(7.5*360/13)}) {$\scriptstyle{\frac{\Nsites+3}{2}}$};
\node[draw=none,font=\tiny,text=black,scale=1.5] at ({2*cos(8.5*360/13)},{2*sin(8.5*360/13)}) {$\cdots$};
\node[draw=none,font=\tiny,text=black,scale=1.5] at ({2*cos(-0.5*360/13)},{2*sin(-0.5*360/13)}) {$\Nsites$};
\node[draw=none,font=\tiny,text=black,scale=1.5] at ({2*cos(-1.5*360/13)},{2*sin(-1.5*360/13)}) {$\scriptstyle{\Nsites-1}$};
\node[draw=none,font=\tiny,text=black,scale=1.5] at ({2*cos(-2.5*360/13)},{2*sin(-2.5*360/13)}) {$\cdots$};
\node[draw=none,font=\tiny,text=black,scale=1.5] at ({2.25*cos(6*360/13)},{2.25*sin(6*360/13)}) {$J_w'$};
\node[draw=none,font=\tiny,text=black,scale=1.5] at ({2.25*cos(7*360/13)},{2.25*sin(7*360/13)}) {$J_w$};
\node[draw=none,font=\tiny,text=black,scale=1.5] at ({2.25*cos(2*360/13)},{2.25*sin(2*360/13)}) {$J_2$};
\node[draw=none,font=\tiny,text=black,scale=1.5] at ({2.25*cos(1*360/13)},{2.25*sin(1*360/13)}) {$J_1$};
\node[draw=none,font=\tiny,text=black,scale=1.5] at ({2.25*cos(0)},{2.25*sin(0)}) {$\;\scriptstyle{-J_f}$};
\node[draw=none,font=\tiny,text=black,scale=1.5] at ({2.3*cos(-1*360/13)},{2.3*sin(-1*360/13)}) {$J_{\scriptscriptstyle{\Nsites-1}}$};
\node[draw=none,font=\tiny,text=black,scale=1.5] at ({2.3*cos(-2*360/13)},{2.3*sin(-2*360/13)}) {$J_{\scriptscriptstyle{\Nsites-2}}$};
\end{tikzpicture}
\caption{The frustrated Ising ring with $\Nsites$ spins. The antiferromagnetic bond $-J_f$ sits at site $j=\Nsites$; the two weak bonds $J_w$ and $J_w'$ flank the central spin at $j=(\Nsites+1)/2$, with $J_w'=J_w$ in the standard symmetric case. All remaining bonds equal $J$ in the standard model and may take distinct values in the modified instances (see text). The dashed blue line indicates the possible reflection symmetry axis.}
\label{fig:ring_gen}
\end{figure}
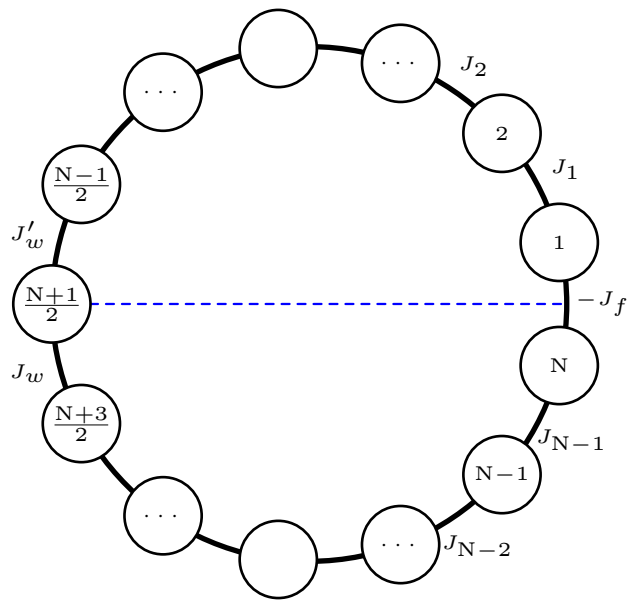

The frustrated Ising ring is a specific instance of $\Ho_z$ in Eq.~\eqref{eqn:H_targ}, defined on $\Nsites$ spins, where $\Nsites$ is odd, with nearest-neighbor couplings and periodic boundary conditions, as illustrated in Fig.~\ref{fig:ring_gen}. All but one of the couplings are ferromagnetic, with the single antiferromagnetic bond introducing frustration — hence the name ``frustrated.'' In the standard case, the couplings $J_j$ take the form:
\begin{equation} \label{eqn:original_ring}
  J_{j} =
    \begin{cases}
      J_w & \text{if $j=(\Nsites\pm 1)/2$}\\
      -J_f & \text{if $j=\Nsites$}\\
      J & \text{otherwise}
    \end{cases}       \;.
\end{equation}
As discussed in \cite{Knysh_PRA2020}, an additional condition is needed for the model to exhibit an exponentially vanishing gap, 
namely $0<J_f<J_w<J$ and $JJ_f>J_w^2$. 
We choose $J_w/J=0.5$, $J_f/J=0.45$ as in \cite{Cote_2023,wang2025exponential}, which satisfy this criterion. 
We also note that the model features a reflection symmetry around the central spin at site $(\Nsites+1)/2$. 
The ground state is the fully ferromagnetic state (all spins aligned), with energy
$E_{\gs} = -(\Nsites-3)J - 2J_w + J_f$. 
The first excited state carries two domain walls
at sites $j=(\Nsites+1)/2$ and $j=\Nsites$, separated from the ground state by an energy
$\Delta_1 = 2(J_w - J_f)$. 
 
As studied in \cite{wang2025exponential,arezzo2025digital} (see Fig.~2 therein), the interpolating Hamiltonian, given in Eq.~\eqref{eqn:ann_hamiltonian_s}, with the frustrated Ising ring as $\Ho_\mathrm{targ}$ exhibits two closing gaps between the ground state and the first excited state during the linear-schedule annealing process: the first  closes polynomially at $s\approx 0.5$ and the second closes exponentially at $s\approx 0.9$. We focus our analysis on the smallest gap because it is the dominant spectral bottleneck for  linear schedules.

The minimum gap for the standard QA interpolating Hamiltonian and the Parity-QA interpolating Hamiltonian $
     s  \tilde{H}_\mathrm{targ} + (1-s  )\tilde{H}_\mathrm{drive} 
$
is given as
\begin{equation}\label{eq:Energy_gap}
\Delta_{\min}= \min_s \Delta_s \hspace{2cm} 
\mathrm{with}\;  \Delta_s = E_d(s) - E_0(s),
\end{equation}
where the integer $d$ represents the ground state's degeneracy of the the target Hamiltonian, while $E_0(s)$ and $E_d(s)$ represent the energies of the ground state and the $d$-th excited state of the interpolating Hamiltonian at a given point of the schedule $s$.

We further introduce variants of the original frustrated ring of Eq.~\eqref{eqn:original_ring} and Fig.~\ref{fig:ring_gen} to support our subsequent analysis and provide richer results and comparisons. 
The first two variants break the reflection symmetry around the central site $j=(\Nsites+1)/2$ and increase the number of distinct coupling values $J_j$, while keeping the property of an exponentially closing gap: 
(i) we set $J_1/J = J_2/J = 0.8$ and keep the remaining couplings unchanged; (ii) Building on (i), we additionally set $J_{\Nsites-1}/J = 0.85$. 
Ref.~\cite{arezzo2025digital} shows that these changes preserve the presence of an exponentially closing energy gap and hence the difficulty for linear-schedule QA. These variants are useful to investigate correlations between the performance of Parity-QAOA and the number of distinct coupling values appearing in the target Hamiltonian (see Sec.~\ref{sec:improved controllability}).

The third variant is an instance of the frustrated ring of Eq.~\eqref{eqn:original_ring}, which is inspired by Refs.~\cite{Maric_2020,Maric_CommPhys2020,Torre_2021}: it consists of setting $J_w/J = 1$ and $-J_f/J=-1$, so that all couplings have unit magnitude and the ring carries a single antiferromagnetic bond of equal magnitude with respect to all other ferromagnetic couplings. 
The product of all $J_j$ is $\prod_j \frac{J_j}{J}=-1$, so no spin configuration can satisfy all bonds simultaneously: exactly one bond must host a domain wall, which can sit on any of the $\Nsites$ bonds. 
The classical ground space is therefore $2\Nsites$-fold degenerate ($\Nsites$ domain-wall positions times the global spin flip), in contrast to the two-fold degenerate ground state of the original frustrated ring of Eq.~\eqref{eqn:original_ring}.

\section{Results} \label{sec:results}
First, we present the minimal energy gap $\Delta_{\min}$, Eq.~\eqref{eq:Energy_gap}, obtained from standard-QA and Parity-QA simulations. We then turn to QAOA simulations and examine whether the trends observed in QA carry over to QAOA.
Simulation details of all simulations can be found in App.~\ref{sec:app:simulation_details}.
\subsection{Minimal Energy gap in QA}\label{sec:min_gap}
\begin{figure}[htp]
\centering
\includegraphics[width=\columnwidth]{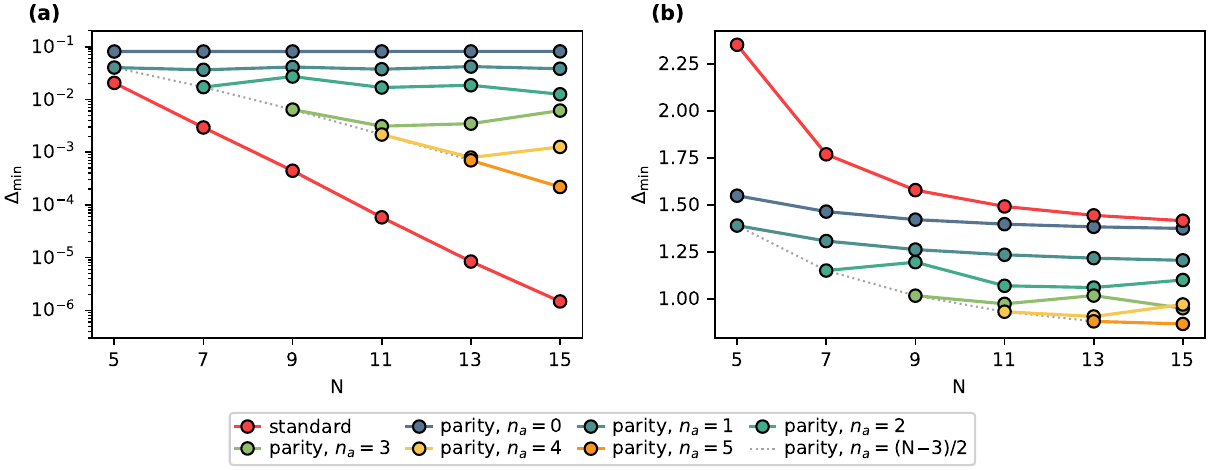}
\caption{Minimal energy gap $\Delta_{\min}$ versus system size $\Nsites$ in the standard and parity implementations, for (a) the frustrated ring model (constraint strength $C=J$; note the logarithmic scale) and (b) the ring with uniform coupling magnitudes and a single antiferromagnetic bond (constraint strength $C=2J$; linear scale). In both panels, solid coloured lines show the parity gap at fixed ancilla number $n_a = 0, 1, \dots, 5$, each starting at the smallest admissible ring $\Nsites = 2n_a + 3$; the grey dotted line is the realistic parity gap at the hardware-required $n_a = (\Nsites-3)/2$. In (a), the standard and realistic-parity gaps close as $\sim e^{-0.96\,\Nsites}$ and $\sim e^{-0.51\,\Nsites}$, so the parity mapping roughly halves the decay exponent. In (b), no exponential closing occurs: all gaps saturate to finite values, with each additional ancilla lowering the gap only moderately (realistic parity: from $1.39$ at $\Nsites=5$ to $0.88$ at $\Nsites=13$).}
\label{fig_notes:MinGap}
\end{figure}

The frustrated Ising ring model in its original formulation, Eq.~\eqref{eqn:original_ring} (and the first two introduced variants in Sec.~\ref{sec:models}) has an exponential closing gap. This is verified in Fig.~\ref{fig_notes:MinGap} (a) where the red line, labeled as `standard QA', shows the results for $\Delta_{\min}$ for $\Nsites = 5,7,\cdots,15$.
The figure also shows the scaling for Parity-QA's interpolating Hamiltonian, constraint strength set to $C=J$, for different numbers of ancilla qubits $n_a$.

As discussed in Sec.~\ref{sec:parity_mapping}, ancilla qubits are needed to break the `unrealizable' K-body constraint into smaller ones. Figure~\ref{fig_notes:mapping} presents an example for $\Nsites=7$, where the dashed colored lines (red and blue) in the logical graph (a) represent auxiliary interactions between qubits that do not interact in the original model. These interactions are transformed into the red and blue ancilla physical qubits in panel (d). 
There are proposals for QA hardware based on superconducting qubits~\cite{Miyazaki_PRA2025, Stenzel_PRA2026} and rydberg atoms ~\cite{Glaetzle_NatComm2017, Lanthaler_PRL2023}  with the latter being capable, in principle, of realizing the configuration shown in Fig.~\ref{fig_notes:mapping} (d) on currently available hardware.

Figure \ref{fig_notes:MinGap}(a) shows different scalings of the gap in the parity framework. When $n_a=0$, there is a clear amplification of the minimal gap, visible even at a small system size $\Nsites=5$.  The gap then remains roughly unchanged with the system size and no exponentially closing behavior is observed. Therefore, using Parity-QA to solve the frustrated ring model would in principle avoid the bottleneck associated with the exponentially closing gap. However, implementing the Hamiltonian without ancilla qubits remains beyond current hardware capabilities, as it requires a global constraint acting across all qubits. To be implementable on annealing hardware, $(\Nsites-3)/2$ ancilla qubits are needed for $\Nsites$ logical qubits. 
The black dotted line in Fig. \ref{fig_notes:MinGap}(a) demonstrates the scaling of the minimal gap when the required number of ancilla qubits is used. 
With this more realistic consideration, the scaling of the spectral gap is back to an exponential decay, albeit with a smaller exponent: log-linear fits over $\Nsites=5,\dots,15$ give $\Delta_{\min}/J\sim e^{-0.96\Nsites}$ for standard QA and $\Delta_{\min}/J\sim e^{-0.51\Nsites}$ for the realistic parity implementation, so the parity mapping roughly halves the decay exponent.

The third variant introduced in Sec.~\ref{sec:models} ($J_j/J\in\{-1, 1\}$) has no exponential closing gap.
This is shown in Fig.~\ref{fig_notes:MinGap}(b). For both standard QA and ancilla-free Parity-QA ($n_a=0$), the gap appears to saturate to a finite value. When ancillas are added ($n_a>0$), the gap decreases but does not close exponentially and again appears to saturate. For this coupling choice, Parity-QA has a smaller gap than for the original frustrated ring; see Fig.~\ref{fig_notes:MinGap}(a). This model poses no small-gap bottleneck for standard QA or Parity QA. Nevertheless, it is instrumental for understanding the role of the constraint unitary in Parity-QAOA, as we elaborate in Sec.~\ref{sec:use_of_constraint}.  

We now turn to analyzing the performance of Parity-QAOA on these models.

\subsection{Parity-QAOA} 
We first show simulation results of the original frustrated Ising ring and its first two variants outlined in Sec.~\ref{sec:models}, before showing results of the third variant. Simulation details can be found in App.~\ref{sec:app:simulation_details}.

\subsubsection{Improved Controllability}\label{sec:improved controllability}
Parity-QAOA simulations were performed on the original frustrated ring and on the modified cases obtained by (i) setting $J_1=J_2=0.8$ and (ii) further setting $J_{\Nsites-1}=0.85$; see Sec.~\ref{sec:models}. There are three distinct $J_j$ values in the original case, four in case (i), and five in case (ii).

\begin{figure}[htp]
\includegraphics[width=0.6\columnwidth]{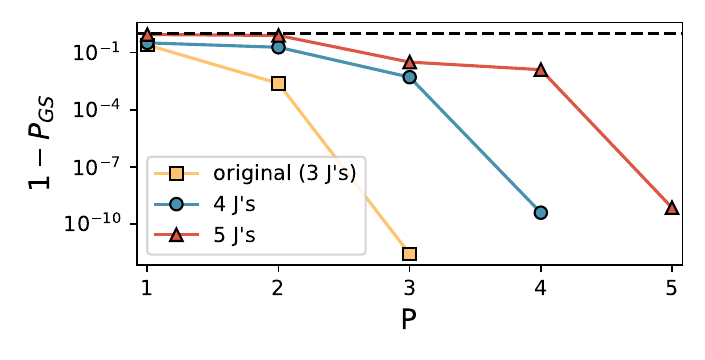}
\caption{Ground state infidelity $1-P_{\mathrm{GS}}$ vs circuit depth $\Ptrot$ (only single qubit z- and x-rotations) for the original formulation of the frustrated ring model and its adaptations for $\Nsites=7$. The greater the number of distinct $J_j$ values, the more layers $\Ptrot$ are needed to reach the ground state.}
\label{fig_notes:Results_OriginalRing}
\end{figure}

For \(\Nsites=7\), the ground-state probabilities are summarized in Fig.~\ref{fig_notes:Results_OriginalRing}.  
Our simulation results show that the ground state is consistently obtained, with unit probability $P_{\mathrm{GS}} = | \langle \tilde{\psi}_{\mathrm{GS}} |\tilde{\psi}_{\Ptrot}(\btheta^x,\btheta^c,\btheta^z)\rangle |^2=1$, at circuit depth $\Ptrot$ equal to the number of distinct $J_j$ values appearing in the Hamiltonian.

Furthermore, we observe that $\theta^c_p=0$, for $p=1, \cdots \Ptrot$, which means that Parity-QAOA reaches the exact ground state of the frustrated ring and its variants \emph{without} requiring the constraint unitary \(\tilde{U}_c(\theta^c_p)\):
It suffices to alternate applications of the local-field unitary \(\tilde{U}_z(\theta^z_p)\), corresponding to single-qubit \(z\)-rotations, and the mixing unitary \(\tilde{U}_x(\theta^x_p)\), corresponding to single-qubit \(x\)-rotations. 
This is a stark contrast to Parity-QA, where it is essential to include a non-zero constraint term $\tilde{H}_c$ in the target Hamiltonian $\tilde{H}_\mathrm{targ}$ to ensure a valid outcome.

The fact that Parity-QAOA attains the ground state using only single-qubit rotations provides direct evidence that its performance is not governed by the spectral gap structure of the interpolating Hamiltonian. Rather, analogously to standard QAOA~\cite{arezzo2025digital}, the success of Parity-QAOA is determined by the controllability of the digitized dynamics and the structure of the reachable manifold, rather than by adiabatic considerations. Consistent with this, the same number of QAOA layers and the same set of optimized parameters can solve the problem across all system sizes $\Nsites$ considered. The following sections outline this behavior and explain why Parity-QAOA can succeed without the constraint unitary $U_c$.

\subsubsection{Optimal control parameters and transferability} \label{sec:parameters}
We observe for small system sizes, $\Nsites=7$, that Parity-QAOA consistently selects trivial constraint unitaries, with $\theta^c_p=0$, for the frustrated Ising model. We can classically simulate the single-qubit rotations $\tilde{U}_x(\theta^x_p)$ and  $\tilde{U}_z(\theta^z_p)$ to obtain a set of optimized parameters $(\btheta^x,\btheta^z)$ that drive the system to the exact ground state, independent of system size. 
The resulting optimal parameters are listed in Table~\ref{tab:SingleQubitRotation}.
The analytic, system-size-independent structure that makes this single-qubit optimization possible is derived in Appendix~\ref{sec:app:derivation_groundstateprobbility}.

\begin{table*}
\begin{center}
\begin{tabular}{ c c c c c c c c c c}
 $\theta^z_1$ &  $\theta^x_1$ &  $\theta^z_2$ &  $\theta^x_2$ &  $\theta^z_3$ &  $\theta^x_3$ & $\theta^z_4$ &  $\theta^x_4$ &  $\theta^z_5$ &  $\theta^x_5$ \\
 \hline\hline
-9.60934341 & 0.20952926 & -8.08672464 & 0.10182431 & -5.23233597 & 0.0988746 & -& -&- &-  \\  
  -5.83795276 & 0.17247357 & 4.92515321 & 0.1598907 & 9.21350412 & 0.10889695 & 4.23400112 &   0.12517238 & - & - \\
  -0.66065018 & 0.18758133 & -2.3049883  & 0.19599091 & 4.2849489 & 0.17213241 & 1.27555834 & 0.21802607 & 6.45839022 & 0.13112482
\end{tabular}
\caption{Parameter values to find the ground state for three different problem cases, in units of $\pi$. The first row shows the results for the original problem formulation  in Eq.~\eqref{eqn:original_ring}. The second row shows results for the modified ring with  $J_1=J_2=0.8$. 
For the third row, we further modified $J_{\Nsites-1}=0.85$.}
\label{tab:SingleQubitRotation}
\end{center}
\end{table*}

\begin{figure}[htp]
\includegraphics[width=0.95\columnwidth]{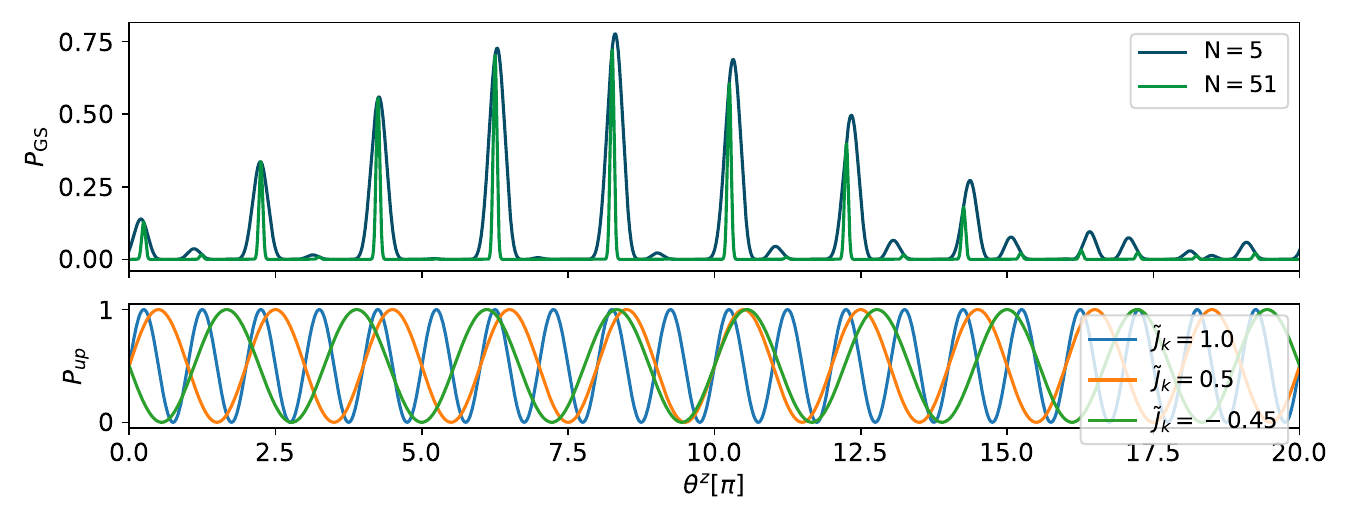}
\caption{The probability of being in the exact ground state $P_{\mathrm{GS}}$ (upper plot) and the probability of a single physical qubit being in the up state $P_{\uparrow}$ (lower plot) as a function of the single control parameter $\theta^z$ at fixed $\theta^x=\pi/4$ (a single QAOA layer, $\Ptrot=1$). In the upper plot we show comparisons between $\Nsites=5$ and $\Nsites=51$. In the lower plot we show comparisons of qubits associated with different parameters $\tilde{J}_k=1.0, 0.5$ and $-0.45$.}
\label{fig_notes:Subplot_Fidelity_Gstart0Gend20}\label{fig_notes:groundstate_probability_p1}
\end{figure}

Consider a single layer ($\Ptrot=1$) of the Parity-QAOA ansatz. Since the constraint unitary is omitted here ($\theta^c=0$), the layer consists only of $\tilde{U}_z(\theta^z)$ and $\tilde{U}_x(\theta^x)$, which act on each parity qubit separately and do not couple different qubits. Each parity qubit $k$ therefore evolves on its own under $\nep^{i \theta^x_p \tilde{\sigma}^x_k}$ and 
$\nep^{i \tilde{J}_k\theta^z_p \tilde{\sigma}^z_k}$
from the initial state $|+\rangle$, and, as the ground state is the all-up configuration, $\ket{\uparrow}^{\otimes K}$, the ground-state probability factorizes into single-qubit up-probabilities,
\begin{equation} \label{eqn:PGS_factorized_main}
P_{\mathrm{GS}}(\theta^x,\theta^z)=\prod_{k=1}^K P_\uparrow(\theta^x,\tilde{J}_k\theta^z) \;,
\end{equation}
with $P_\uparrow(\theta^x,\tilde{J}_k\theta^z)=|\langle\uparrow \!|\nep^{i \theta^x_p \tilde{\sigma}^x_k}\nep^{i \tilde{J}_k\theta^z_p \tilde{\sigma}^z_k}
|+\rangle|^2$ the probability that qubit $k$ is in the ${\ket{\uparrow}}$ state after one layer; the derivation is given in Appendix~\ref{sec:app:derivation_groundstateprobbility}.
Figure~\ref{fig_notes:groundstate_probability_p1} evaluates these single-layer expressions with $\theta^x=\pi/4$ held fixed and the single parameter $\theta^z$ scanned (in units of $\pi$): the lower panel shows the single-qubit factor $P_\uparrow$ for each of the three distinct couplings $\tilde{J}_k=1,\,0.5,\,-0.45$ of the original ring, and the upper panel their product $P_{\mathrm{GS}}$ for $\Nsites=5$ and $\Nsites=51$.

Because the three couplings act as different effective rotation rates, their single-qubit factors peak at different values of $\theta^z$, so a single layer cannot bring all qubits to the ground state simultaneously and the depth-one $P_{\mathrm{GS}}$ remains below one. The product is large only where the factors are simultaneously large, and is suppressed where they are not. 
We find that reaching $P_{\mathrm{GS}}=1$ requires a number of layers $\Ptrot$ equal to the number of distinct couplings, three here, which we report in Table~\ref{tab:SingleQubitRotation}.
The product probability peaks of $P_{\mathrm{GS}}$ also sharpen markedly from $\Nsites=5$ to $\Nsites=51$, indicating that larger systems require more finely tuned parameters;
this suggests that when transferring parameters optimized for a small $\Nsites$, additional refinement is needed. 
In particular, transferring parameters from smaller to larger $\Nsites$ seems necessary to navigate barren plateaus, which become more pronounced for larger $\Nsites$~\cite{McClean_NatureCommunications2018}.

This analysis shows that, to reach the ground state, qubits with the same $\tilde J_k$ should behave the same and there is no need to separate them, while individual control of qubits is needed when the $\tilde J_k$ are different. In fact, the whole problem can be reduced to qubits that equal the number of distinct couplings $\tilde J_k$ in the problem.
Here, we have three distinct $\tilde J_k$ and each corresponding qubit accumulates a different phase under $U_z$, as shown in the bottom panel of Fig.\ref{fig_notes:groundstate_probability_p1}. Given certain properties of the $\tilde J_k$ this frequency selectivity allows one to isolate actions on each qubit and, therefore, allow single qubit control. The proof in App.~\ref{sec:app:local_controllability} shows that the needed properties for arbitrary single qubit control are that the given $\tilde J_k$ and their absolute value are distinct, $|\tilde J_k| \neq |\tilde J_l|$ for any $k\neq l$.
Hence, when $\tilde J_{k}\in\{-1,1\}$ the qubits with $\tilde J=1$ and $\tilde J'=-1$ cannot be separated with single-qubit rotations only and Parity-QAOA must include $\tilde{U}_c$ in this setting ($\theta^c_p \neq 0$). 

\subsubsection{Use of Constraints}
\label{sec:use_of_constraint}
We now turn to the regime in which the constraint unitary is indispensable. As described in Sec.~\ref{sec:models}, we set all coupling magnitudes to unity: $J_j/J = 1$ for all couplings except for a single coupling, $J_f/J = -1$. 
In the parity representation, the ground state is characterized by all qubits aligning with their local fields $\tilde{J}_k$, except for a single qubit, which must oppose its local field.
As shown in App.~\ref{sec:app:local_controllability}, when there is $J'=-J$ the qubit with the negative local field $J'$ will always end up in the opposite state than the other one by only using a sequence of $\tilde{U}_z$ and $\tilde{U}_x$ alone.
Therefore, the constraint-free mechanism identified in Sec.~\ref{sec:improved controllability} cannot align all qubits with their ground-state orientations, making it necessary to actively employ $\tilde{U}_c$. 
Note also that frustration makes the ground space degenerate: the single unsatisfied bond can sit on any of the $\Nsites$ bonds, yielding a $2\Nsites$-fold degenerate logical ground manifold ($\Nsites$-fold in the parity representation). 
Throughout this section, we therefore define $P_{\mathrm{GS}} = \langle \tilde{\psi}_\Ptrot | \hat{\mathcal{P}}_{\mathrm{GS}} | \tilde{\psi}_\Ptrot \rangle$, 
with $\hat{\mathcal{P}}_{\mathrm{GS}}$ the projector onto the ground manifold.

\begin{figure*}[htp]
\includegraphics[width=\textwidth]{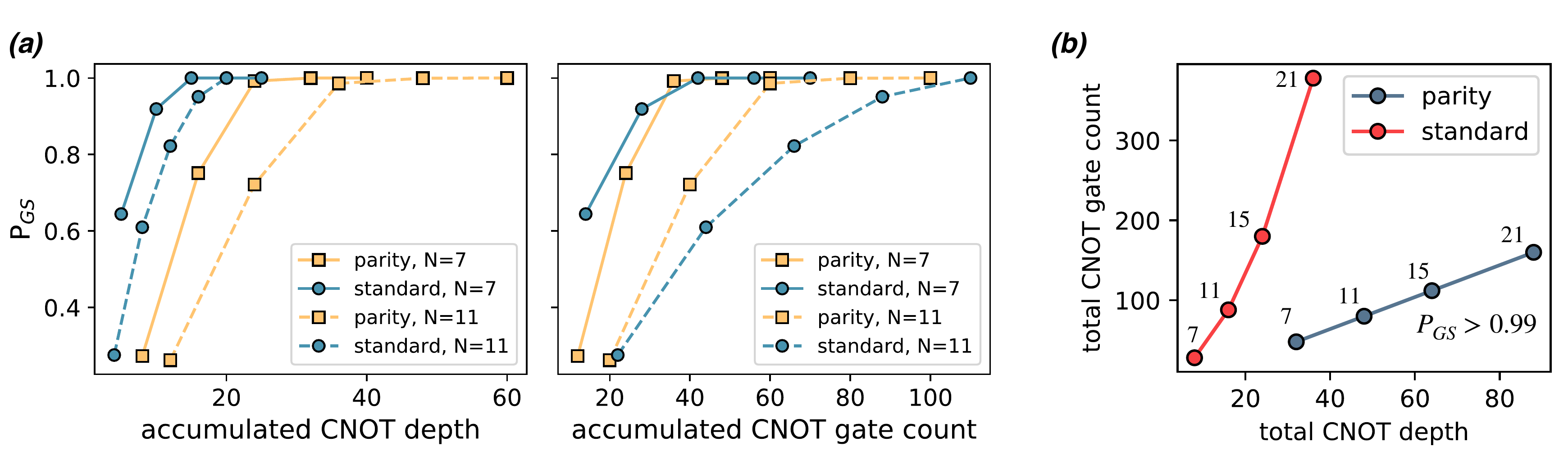}
\caption{Performance and resource scaling of standard and Parity-QAOA on the modified frustrated ring ($-J_f/J=-1$). (a)~Ground-state probability $P_{\mathrm{GS}}$ as a function of the accumulated CNOT depth (left) and CNOT gate count (right) for $\Nsites = 7$ and $11$; successive markers correspond to $\Ptrot = 1, 2, \ldots, 5$. At fixed $\Ptrot$, $P_{\mathrm{GS}}$ decreases significantly with $\Nsites$ for standard QAOA but not for Parity-QAOA. (b)~Total CNOT depth and CNOT gate count required to reach $P_{\mathrm{GS}} > 0.99$, shown for $\Nsites = 7, 11, 15, 21$. Standard QAOA needs a total depth of $\mathcal{O}(\Nsites)$ and a total gate count of $\mathcal{O}(\Nsites^2)$. Parity-QAOA needs a total depth and a total gate count that both scale as $\mathcal{O}(\Nsites)$.}
\label{fig_notes:Results_SingleRing}
\end{figure*}

The numerical results, summarized in Fig.~\ref{fig_notes:Results_SingleRing}
(a), show $P_{\mathrm{GS}}$ displayed as a function of the accumulated CNOT depth (left) and CNOT gate count (right), with successive markers corresponding to $\Ptrot = 1, 2, \ldots$. For standard QAOA, $P_{\mathrm{GS}}$ at fixed $\Ptrot$ decreases with increasing $\Nsites$, and the exact ground state ($P_{\mathrm{GS}} = 1$) is first reached at $\Ptrot^* = \lceil \Nsites/2 \rceil - 1$ (i.e., $\Ptrot^* = 3, 5, 7, 10$ for $\Nsites = 7, 11, 15, 21$). 
For Parity-QAOA, by contrast, 
$P_{\mathrm{GS}} > 0.99$ is reached at a constant $\Ptrot^* = 4$ for all system sizes studied. The exact ground state is not attained at the depths we simulated (up to $\Ptrot = 5$), but the residual error keeps decreasing with depth (e.g., $1 - P_{\mathrm{GS}} \approx 2 \times 10^{-6}$ at $\Ptrot = 5$ for $\Nsites = 11$), indicating that $P_{\mathrm{GS}} \approx 0.99$ at $\Ptrot = 4$ reflects the chosen depth rather than a limitation of the Ansatz. 
The optimal parameters transfer across system sizes: using the $\Nsites = 11$, $\Ptrot = 4$ optimum as initial guess for a local re-optimization, we verified that $P_{\mathrm{GS}} > 0.99$ is consistently achieved till $\Nsites=21$. 
This can also be done for standard QAOA, see App.~\ref{sec:app:optimal_parameters_standardQAOA}.

We now combine $\Ptrot^*$ with the per-layer costs established in Sec.~\ref{sec:Parity-QAOA} (standard QAOA: $2\Nsites$ CNOTs and constant depth per layer; Parity-QAOA: $2(\Nsites-1)$ CNOTs and depth $\mathcal{O}(\Nsites)$ per layer). Figure~\ref{fig_notes:Results_SingleRing}(b) compares the total resources required to reach $P_{\mathrm{GS}} > 0.99$ at $\Nsites = 7, 11, 15, 21$. 

For standard QAOA, the $\mathcal{O}(\Nsites)$ layers yield a total CNOT depth of $\mathcal{O}(\Nsites)$ and a total CNOT count of $\mathcal{O}(\Nsites^2)$. 
For Parity-QAOA, the constant number of layers yields a total depth that is also $\mathcal{O}(\Nsites)$, with a prefactor approximately twice that of standard QAOA, but a total CNOT count of only $\mathcal{O}(\Nsites)$. On NISQ hardware, where accumulated two-qubit gate errors typically dominate over coherence-time limitations, the total gate count is often the more relevant figure of merit, making this trade-off favorable for Parity-QAOA despite its depth overhead and the smaller residual $1 - P_{\mathrm{GS}}$.

\begin{figure}[htp]
\includegraphics[width=0.6\columnwidth]{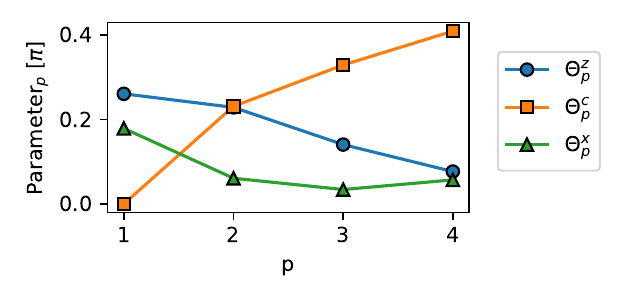}
\caption{Best found Parity-QAOA parameters $(\theta^z_p, \theta^c_p, \theta^x_p)$ at each layer $p=1,\dots,\Ptrot$, for $\Nsites=11$ and $\Ptrot=4$, for the ring with uniform coupling magnitudes and a single antiferromagnetic bond. The label $[\pi]$ on the vertical axis indicates that all angles are expressed in units of $\pi$: a plotted value of $0.2$ corresponds to a rotation angle of $0.2\pi$. 
The constraint unitary is actively engaged, in contrast to the constraint-free regime of Sec.~\ref{sec:improved controllability}; the first constraint angle is trivial, $\theta^c_1 = 0$.}
\label{fig_notes:app:Best_Parameters_Parity_SignedRing}
\end{figure}

The optimal Parity-QAOA parameters at $\Nsites = 11$, $\Ptrot = 4$ are shown in Fig.~\ref{fig_notes:app:Best_Parameters_Parity_SignedRing}: the constraint unitary is actively engaged at the optimum, in direct contrast to the original frustrated ring of Sec.~\ref{sec:improved controllability}. Notably, an optimum with $\theta^c_1 = 0$ exists, so that only three of the four constraint parameters are non-trivial.

\section{Conclusions}
The frustrated Ising ring is one of the simplest models exhibiting an exponentially closing spectral gap, making it a paradigmatic and stringent benchmark for quantum optimization. 
Its parity-compiled variant is particularly well-suited for this study because the ring topology incurs zero qubit overhead and requires fewer constraint terms than the previously studied dense graphs, placing it in a qualitatively different regime of the parity architecture. 
In this work, we have investigated the performance of the parity (LHZ) architecture on this model and its extensions, both in continuous-time quantum annealing and in digitized protocols based on QAOA.

In the continuous-time QA setting, we have shown that the parity mapping enlarges the minimal spectral gap relative to standard linear-schedule QA. 
This advantage is quantifiably reduced when hardware-compatible ancilla qubits are introduced to decompose the global constraint into shorter loops: exponentially closing of the gap is restored, but with a smaller exponent. The result provides an honest, quantified picture of the spectral benefit versus hardware cost trade-off inherent to parity-based annealing on sparse graphs.

Parity--QAOA on the original frustrated ring and its symmetry-broken variants revealed an unexpected regime of \emph{constraint-free} operation. 
The exact ground state is reached using only single-qubit rotations, with circuit depth equal to the number of distinct coupling values in the Hamiltonian and independent of system size.  
To our knowledge, this is the first explicit instance in which the constraint unitary in Parity--QAOA is entirely unnecessary. The mechanism --- qubits with distinct local fields in the parity basis have arbitrary qubit control --- is concretely understood and is unlikely to work in more complex problem settings.

To probe a case in which Parity-QAOA's constraint unitaries are necessary to prepare the ground state,  we further considered a variant of the model where all coupling magnitudes are set to unity, with a single antiferromagnetic bond. 
In this regime, Parity-QAOA achieves $P_\mathrm{GS} \approx 0.99$ with a number of layers that remains constant over the system sizes studied, while reducing the total CNOT count from $\mathcal{O}(\Nsites^2)$ to $\mathcal{O}(\Nsites)$ with respect to standard QAOA. 
The total circuit depth is $\mathcal{O}(\Nsites)$ in both approaches but Parity-QAOA exhibits a factor of two depth overhead. Interestingly, Ref.~\cite{weidinger2024performance} found the opposite trade-off for complete graphs, where Parity-QAOA requires more two-qubit gates than standard QAOA  but offers advantages in circuit depth.
Taken together with the present results, this suggests that standard QAOA and Parity-QAOA are two complementary approaches, whose relative resource requirements  strongly depend on the density of the underlying problem graph.
This observation motivates extending the present analysis beyond rings to broader classes of sparse graphs, where the structure of the parity constraints and the associated ancilla requirements may also lead to qualitatively different scaling of circuit depth and two-qubit gate count.

On the experimental side, implementing Parity-QA and Parity-QAOA for these ring instances on current annealing and gate-based hardware would provide a direct test of whether the spectral gap amplification and gate-count reduction identified here survive realistic noise and connectivity constraints.

\vspace{5mm}
A Python code implementation of the QAOA circuits considered in this work is available online \cite{Ring_GitlabTutorial}.

\begin{acknowledgments}
A.W. thanks Julian Farnsteiner and Francensco Ghisoni for useful discussions.
This study was supported by NextGenerationEU via FFG and Quantum Austria (FFG Project No. FO999896208).
This research was funded in part by the Austrian Science Fund (FWF) under Grant-DOI 10.55776/F71 and Grant-DOI 10.55776/Y1067.
This project was funded within the QuantERA II Programme that has received funding from the European Union’s Horizon 2020 research and innovation programme under Grant Agreement No. 101017733. 
This publication has received funding under Horizon Europe programme HORIZON-CL4-2022-QUANTUM-02-SGA via the project 101113690 (PASQuanS2.1).
This study was supported by the Federal Ministry for Economic Affairs and Climate Action through project QuaST.
For the purpose of open access, the author has applied a CC BY public copyright license to any Author Accepted Manuscript version arising from this submission.
\end{acknowledgments}

\appendix
\section{Simulation details}\label{sec:app:simulation_details}
To calculate the minimal Energy gap $\Delta_{\min}$ of Eq.~\eqref{eq:Energy_gap} we need to get the instantaneous energy spectra of the interpolation Hamiltonian. We can do this by simulating the time evolution of the time dependent Hamiltonian using QuTiP \cite{Johansson_QuTiP2012,Johansson_QuTiP2013}.
However, this is only efficient for $\Nsites<14$. 
For bigger systems, to get the low lying eigenenergies, we use the iterative eigensolver (\emph{eigsh}) from \textsc{SciPy} \cite{Virtanen_SciPy2020}. 
Rather than constructing the Hamiltonian as a dense matrix, it was represented as a linear operator. This allows the Lanczos algorithm underlying function \emph{eigsh} to determine the lowest eigenvalues and corresponding eigenvectors using only repeated matrix-vector products, avoiding the storage and manipulation of the full Hamiltonian matrix.

All QAOA simulations are exact state-vector simulations performed with \textsc{Qiskit}~\cite{qiskit2024}.
For all simulations the parameter domains were restricted to 
\(\theta^x_p \in [0, \pi/4]\), and \(\theta^c_p \in [-\pi/2, \pi/2]\). 
For the first variants, the parameter domain for $\theta^z_p$ was set to \(\theta^z_p \in [-10\pi, 10\pi]\) and for the third variant (as $\tilde J_k = |1|$) to \(\theta^z_p \in [-\pi/2, \pi/2]\).
The constraint strengths were set to $C = J$ for the first variants and $C = 2J$ for the third variant where $\tilde J_k\in \{-1, 1\}$.
For each problem instance, we used
100 random restarts 
of the full parameter vectors $(\btheta^x,\btheta^z)$ for standard QAOA and $(\btheta^x,\btheta^z,\btheta^c)$ for Parity-QAOA and optimized the parameters using the \textsc{BFGS} \cite{nocedal1999numerical} routine from python \textsc{SciPy} \cite{Virtanen_SciPy2020}, retaining the best result.

\section{Ground state probability for Parity-QAOA at arbitrary size}
\label{sec:app:derivation_groundstateprobbility}
In the original formulation of the frustrated ring model, Parity-QAOA uses single-qubit rotations only to get to the ground state. 
The needed QAOA parameters can be found in Table~\ref{tab:SingleQubitRotation}. 
They are obtained as follows: Using the best found parameters $({\theta^x}^*, {\theta^z}^*)$ obtained from QAOA simulations  $\Nsites=7$ of QAOA simulations one can use a classical algorithm to refine these parameters (BFGS search) so that the ground state probability $P_{\mathrm{GS}}=1$ also for large $\Nsites>10^6$.
The classical algorithm works the following way:
The unitaries for single qubit rotations are
\begin{equation}
    \tilde{U}_x(\theta^x_p) = \prod_{k=1}^K 
    \nep^{i \theta^x_p \tilde{\sigma}^x_k} = 
    \prod_{k=1}^K \big( \cos{(\theta^x_p)} \identity^k + i \sin{(\theta^x_p}) \tilde{\sigma}^x_k \big) \;,
\end{equation}
\begin{equation}
    \tilde{U}_z(\theta^z_p) = 
    \prod_{k=1}^K  \nep^{i \tilde{J}_k \theta^z_p \tilde{\sigma}^z_k} = 
    \prod_{k=1}^K  \big( \cos{(\tilde{J}_k\theta^z_p)}\identity^k  + i \sin{(\tilde{J}_k\theta^z_p)} \tilde{\sigma}^z_k \big) \;.
\end{equation}
While $\tilde{U}_x(\theta^x_p)$ represents a rotation around the x-axis which acts in the same way on every physical qubit $k$, the unitary $\tilde{U}_z(\theta^z_p)$ is a product of independent rotations around the z-axis which, however, depend on the local field $\tilde{J}_k$ of the physical qubit $k$.
Since all physical qubits are independent, let us concentrate our attention on qubit $k$.
The state for a qubit at site $k$ starting in $|+\rangle_k=(|\!\uparrow\rangle+|\!\downarrow\rangle)/\sqrt{2}$ after $\Ptrot$ layers of single-qubit rotations is
\begin{equation}
|\tilde{\psi}_k (\btheta^x, \tilde{J}_k\btheta^z)\rangle = \nep^{i \theta^x_{\Ptrot} \tilde{\sigma}^x_k} \, 
\nep^{i \tilde{J}_k \theta^z_{\Ptrot} \tilde{\sigma}^z_k} \, \cdots \,
\nep^{i \theta^x_{1} \tilde{\sigma}^x_k} \, 
\nep^{i \tilde{J}_k \theta^z_{1} \tilde{\sigma}^z_k} \, |+\rangle_k \;.
\end{equation}
The ground state has all physical qubits in the up state $|\!\uparrow\rangle$. 
One can calculate the probability that qubit $k$ is in the up state as
$P_{\uparrow}(\btheta^x, \tilde{J}_k \btheta^z) = 
|\langle  \uparrow | \tilde{\psi}_k (\btheta^x, \tilde{J}_k \btheta^z) \rangle |^2$.
The total probability of being in the ground state can be obtained by multiplying $P_{\uparrow}(\btheta^x, \tilde{J}_k \btheta^z)$ for all $K$ qubits:
\begin{equation}
    P_{\mathrm{GS}}(\btheta^x, \btheta^z) = 
    \prod_{k=1}^K P_{\uparrow} (\btheta^x, \tilde{J}_k \btheta^z) \;.
\end{equation}
In the case of the original formulation with three different $J_j$ values ($\Nsites-3$ qubits with $\tilde{J}_k=J$, two qubits with $\tilde{J}_k=J_w$ and one qubit with $\tilde{J}_k=-J_f$), the simplified formula for the ground state probability is
\begin{equation}
    P_{\mathrm{GS}}(\btheta^x, \btheta^z) = P_{\uparrow}^{\Nsites-3}(\btheta^x, J\btheta^z) P_{\uparrow}^2(\btheta^x, J_w\btheta^z) P_{\uparrow}(\btheta^x, -J_f\btheta^z) \;.
    \label{eqn:app:PGS_factorized}
\end{equation}
The aim is to find the parameters $({\btheta^x}^*, {\btheta^z}^*)$ so that $P_{\mathrm{GS}}({\btheta^x}^*, {\btheta^z}^*) = 1$. The optimized parameters $({\btheta^x}^*,{\btheta^z}^*)$ found in this way are exactly those reported in Table~\ref{tab:SingleQubitRotation} (first row, for the original three-$J$ ring), and Eq.~\eqref{eqn:app:PGS_factorized} is precisely what is plotted in Fig.~\ref{fig_notes:groundstate_probability_p1}: the lower panel shows the individual single-qubit probabilities $P_\uparrow(\btheta^x,\tilde{J}_k\btheta^z)$ for $\tilde{J}_k=1,\,0.5,\,-0.45$, and the upper panel their product $P_{\mathrm{GS}}$, so that the exact ground state is reached at parameter values where all three factors are simultaneously maximal.

\section{Local controllability}\label{sec:app:local_controllability}
In the original formulation of the frustrated ring there are three different coupling strengths, $J_1=1$, $J_2=0.5$ and $J_{\Nsites-1}=-0.45$.
In the Parity mapping the couplings become local fields and, as shown in the main text, single qubit rotations are enough to find the ground state (all qubits up). 

To be able to get the ground state one needs to align the qubits with the fields $J_1$ and $J_2$ to their local field, while the qubit with the negative field $J_{\Nsites-1}$ has to oppose its field. 
Therefore, it is necessary to be able to address qubits with different local fields $J_k$ individually. Obviously, it is not possible to address qubits with the same local field separately, and here it is also not necessary as we want qubits with the same field $J_k$ to behave the same.

The following proof is based on the dynamical Lie algebra formalism widely used in optimal quantum control. The key step reduces the controllability condition to the invertibility of a Vandermonde matrix~\cite{Albertini_SciDirect2002}. This proof does not say how many layers of single qubit rotations are needed to achieve individual control, just that it is possible.

Consider the Hamiltonian $H_x = \sum_k \sigma_k^x$ and $H_z = \sum_k^K J_k \sigma_k^z$, where we only use $M$ distinct $J_k$ (in the above example $M=3)$. 
To determine the set of Hamiltonians that can be synthesized from the available controls, we consider the Lie algebra generated by $H_z$ and $H_x$, which is obtained by forming consecutive commutators. First $[H_z, H_x] = 2i \sum_k J_k \sigma_k^y$ and further $[H_z, [H_z, H_x]]= -4 \sum_k J_k^2 \sigma_k^x$. Continuing this yields Hamiltonians proportional to $J_k^m \sigma_k^P$ where $P=x$ when $i$ is even and $P=y$ when $m$ is odd. First, we focus on the even powers $H^{(m-1)}=\sum_k^K J_k^{2m} \sigma_k^x$ (omitting the unnecessary prefactors $(2i)^{2m}$ for simplicity) and write this equation as
\begin{equation}\label{eq:generating_Hamiltonian}
    \begin{pmatrix}
     H^{(0)} \\
     H^{(1)} \\
     \vdots  \\
     H^{(M-1)}
\end{pmatrix}
= V_{even}
    \begin{pmatrix}
     \sigma^x_1\\
     \sigma^x_2 \\
     \vdots  \\
     \sigma_{M}^x
\end{pmatrix},
\end{equation}
where the coefficients form a Vandermonde matrix $V_{even}$ given as
\begin{equation}
    V_{even}= 
    \begin{pmatrix}
     1 & 1 & \cdots & 1\\
     J_1^2 & J_2^2 & \cdots & J_K^2 \\
     \vdots & \vdots & &\vdots \\
     J_1^{(2M-2)} & J_2^{(2M-2)} & \cdots & J_K^{(2M-2)}
\end{pmatrix}.
\end{equation}
Here, $det(V_{even})\neq 0$, as the $J_k$ and their powers $J_k^m$ are distinct, which means that $V_{even}$ is invertible. Similarly, for odd powers we can write the equation  $H^{(m-1)}=\sum_k^K J_k^{2m-1} \sigma_k^y$, where the resulting Vandermonde matrix is
\begin{equation}
    V_{odd}= 
    \begin{pmatrix}
     J_1 & J_2 & \cdots & J_K\\
     J_1^3 & J_2^3 & \cdots & J_K^3 \\
     \vdots & \vdots & &\vdots \\
     J_1^{(2M-1)} & J_2^{(2M-1)} & \cdots & J_K^{(2M-1)}
\end{pmatrix}.
\end{equation}
The matrix $V_{odd}$ can be inverted if $|J_k| \neq |J_l|$ for $k \neq l$. The absolute value is important because when $J_k = -J_l$ the powers $J_k^{(m)}$ are distinct but the columns $k$
 and $l$ only differ by a constant factor $-1$ which makes them dependent and, therefore, the matrix would not be invertible. Therefore, for both matrices $V_{even}$ and  $V_{odd}$ the sufficient condition for invertibility is that $|J_k| \neq |J_l|$ for $k \neq l$.

Using the invertible Vandermonde matrix $V_{even}$ in Eq.~\eqref{eq:generating_Hamiltonian}, we can rewrite it to
\begin{equation}
    \begin{pmatrix}
     \sigma_1^x \\
     \sigma_2^x \\
     \vdots  \\
     \sigma_M^x
     \end{pmatrix}
= V^{-1}_{even}
    \begin{pmatrix}
     H_x^{(0)} \\
     H_x^{(1)} \\
     \vdots  \\
     H_x^{(M-1)}
\end{pmatrix}
\end{equation}
This equations shows that every individual operator $\sigma_k^x$ can be expressed as a linear combination of the generated Hamiltonians. 
 As $[H_z, \sigma_k^x] = 2i J_k\sigma_k^y$ and $[\sigma_k^y, \sigma_k^x] = 2i\sigma_k^z$ the full set $\{\sigma_k^x, \sigma_k^y, \sigma_k^z\}$ can be generated for every qubit $k$, which implies arbitrary single-qubit control. 
This proof also works when considering $J_{\Nsites-1}=0.8$ and $J_4=0.85$ in addition to the previous mentioned $J_k$ as all $|J_k|$ are different.

As mentioned, in the case of opposite signs of the local fields $J_1=1$ and $J_2=-J_1$ the above proof fails as $V$ cannot be inverted. 
In fact, one can show that qubits with the negative local field $J_2$ will always be flipped compared to the ones with the positive field $J_1$ if only single qubit rotations are used. 
This can be proved the following way: 
Consider, for simplicity, only two qubit $1$ and $2$ and the following possible single qubit rotations: a x-rotation $R_x(\theta^x)= e^{-i \theta^x \sigma^x}$ that is the same for both qubits, and a z-rotation that is $R_z(\theta^z) = e^{-i \theta^z \sigma^z}$  for qubit $1$ and $R_z(-\theta^z)$ for qubit $2$ (because of the opposite signs of their local fields $J_k$). 

Assume, qubit 1 experiences the following $p$ sequences of single qubit rotations

\begin{equation}\label{eq:app:qubit_one}
    \ket{\psi}_1 = R_x(\theta^x_p) R_z(\theta^z_p) \dots R_x(\theta^x_1) R_z(\theta^z_1) \ket{+}_1 .
\end{equation}
For qubit 2 the same sequence leads to the following variational state
\begin{equation}
    \ket{\psi}_2 = R_x(\theta^x_p) R_z(-\theta^z_p) \dots R_x(\theta^x_1) R_z(-\theta^z_1) \ket{+}_2 .
\end{equation}

One can exploit the fact that $\sigma^x \sigma^z \sigma^x = -\sigma^z$ to make the transformation $R_z(-\theta^z) = \sigma^x R_z(\theta^z)  \sigma^x$. Also, $R_x(\theta^x) = \sigma^x R_x(\theta^x)  \sigma^x$. Substituting this into the previous equation gives

\begin{equation}
    \ket{\psi}_2 = (\sigma^x R_x(\theta^x_p)  \sigma^x) (\sigma^x R_z(\theta^z_p)  \sigma^x) \dots (\sigma^x R_x(\theta^x_1)  \sigma^x) (\sigma^x R_z(\theta^z_1)  \sigma^x) \ket{+}_2 .
\end{equation}

Neighboring $\sigma^x$ cancel as $\sigma^x \sigma^x=\identity$, which leaves $\sigma^x$ only at the start and end of the sequence. Considering that $\sigma^x \ket{+} = \ket{+}$ also removes the $\sigma^x$ at the start. Hence, the equation further simplifies to

\begin{equation}\label{eq:app:qubit_two}
    \ket{\psi}_2 = \sigma^x R_x(\theta^x_p) R_z(\theta^z_p)  \dots R_x(\theta^x_1)  R_z(\theta^z_1) \ket{+}_2 .
\end{equation}

Comparing this variational state, Eq.~\eqref{eq:app:qubit_two}, with the variational state of qubit 1, Eq.~\eqref{eq:app:qubit_one}, we see that they differ by one $\sigma^x$ in the end of the single qubit rotation sequence. This means that qubit $2$ is always flipped compared to qubit $1$ - it is not possible to separate the qubits and address them individually.
Hence, if the local fields are opposite, no sequence of single-qubit rotations can rotate both qubits into the same final state.

\section{Optimal parameters for standard QAOA}\label{sec:app:optimal_parameters_standardQAOA}
\begin{figure}[htp]
\includegraphics[width=0.6\columnwidth]{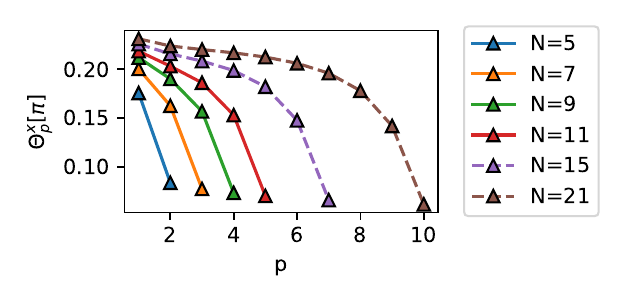}
\caption{Best found QAOA parameters $\theta^x_p$ at each layer $p=1,\dots,\Ptrot$, for varying $\Nsites$ to achieve $P_{GS}>0.99$ for the ring with uniform coupling magnitudes and a single antiferromagnetic bond. Parameters for $\Nsites=5-11$ (solid line) are obtained by taking the best of several optimization runs, parameters for $\Nsites=15, 21$ (dashed lines) are obtained from extrapolation and local optimization. The corresponding parameters $\theta^z_p$ follow the inverse order of $\theta^x_p$.}
\label{fig_notes:app:Best_Parameters_Standard_SignedRing}
\end{figure}

In the main text, Sec.~\ref{sec:use_of_constraint}, simulation results for the ring with uniform couplings $-1, 1$ were presented. The results showed that standard QAOA can find the solution, 
$P_{\mathrm{GS}}=1$, when $\Ptrot^* = \lceil \Nsites/2 \rceil - 1$. Further, we observed that the values of $\theta^z_p$ are related to those of $\theta^x_p$. In particular, $\theta^x_i=\theta^z_j$ for $i=1, 2, \dots P^*$ and $j=P^*, P^*-1, \dots, 1$ ($\theta^z_p$ follows the inverse order of $\theta^x_p$).
For simplicity, we show the best parameters $\theta^x_p$ found in the simulations ($N=5-11$) in Fig.~\ref{fig_notes:app:Best_Parameters_Standard_SignedRing}. The pattern can be exploited to extrapolate to the parameters needed for $N=15, 21$. Refining the results with local optimization (using \textsc{Qiskit}), as described in the main text Sec.~\ref{sec:use_of_constraint}, one can guarantee $P_{\mathrm{GS}}>0.99$. Those parameters are displayed in the Figure with dashed lines. 

\bibliography{biblio}

\end{document}